\documentclass[usenatbib]{mn2e}
\usepackage{natbib}
\bibpunct{(}{)}{;}{a}{,}{,}
\input epsf
\def\plotone#1{\centering \leavevmode
\epsfxsize= 1.0\columnwidth \epsfbox{#1}}

\newcommand{\lesssim}{\raisebox{-.1ex}{\renewcommand{\arraystretch}{0.3}
$\begin{array}{@{}c} \scriptstyle < \\ \scriptstyle \sim \end{array}$}}

\newcommand{\gtrsim}{\raisebox{-.1ex}{\renewcommand{\arraystretch}{0.3}
$\begin{array}{@{}c} \scriptstyle > \\ \scriptstyle \sim \end{array}$}}



\def\vecd{{\mbox{\boldmath $d$}}}
\def\vecx{{\mbox{\boldmath $x$}}}

\def\vecz{{\mbox{\boldmath $z$}}}

\def\veck{{\mbox{\boldmath $k$}}}
\def\vecv{{\mbox{\boldmath $v$}}}
\def\vecr{{\mbox{\boldmath $r$}}}
\def\veceta{{\mbox{\boldmath $\eta$}}}

\long\def\comment#1{}

\def\para{\parallel}

\def\rms{{\em rms}\ }
\def\ie{{i.e.}}
\def\eg{{e.g.}}
\def\etc{{etc.}}

\def\W2{{\cal W}}

\def\del{\nabla}

\def\be{\begin{equation}}
\def\ee{\end{equation}}
\def\bea{\begin{eqnarray}}
\def\eea{\end{eqnarray}}

\def\Mpc{\,{\rm Mpc}}

\def\cmm2{{\,\rm cm^{-2}}}
\def\cm2{{\,{\rm cm}^2}}
\def\cmm3{{\,{\rm cm}^{-3}}}
\def\gcmm3{{\,{\rm g\,cm^{-3}}}}
\def\kms{\,{\rm km\,s^{-1}}}

\def\fun#1#2{\lower3.6pt\vbox{\baselineskip0pt\lineskip.9pt
  \ialign{$\mathsurround=0pt#1\hfil##\hfil$\crcr#2\crcr\sim\crcr}}}

\title[Mass Selection Bias in Galaxy Cluster
  Peculiar Velocities from kSZ]{Mass Selection Bias in Galaxy Cluster
  Peculiar Velocities  from the Kinetic Sunyaev--Zel'dovich Effect}

\author[A. C. Peel]{\bf Alan C. Peel$^{1}$\thanks{E-mail: 
a.peel@damtp.cam.ac.uk}\\
$^{1}$Department of Applied Mathematics and Theoretical Physics,
University of Cambridge, Cambridge CB3 0WA, United Kingdom}
\begin{document}

\date{submitted to MNRAS}

\pagerange{0--0} \pubyear{2005}

\maketitle

\label{firstpage}

\begin{abstract}
Upcoming surveys for galaxy clusters using the Sunyaev--Zel'dovich
effect are potentially sensitive enough to create a peculiar velocity
catalog.  The statistics of these peculiar velocities 
are sensitive to cosmological parameters.  We develop a method to
explore parameter space using N-body simulations in order to
quantify dark matter halo velocity statistics which will be useful 
for cluster peculiar velocity observations.  We show that
mass selection bias from a kinetic Sunyaev--Zel'dovich 
velocity catalog forecasts \rms peculiar velocities with a much more
complicated 
$\Omega_m$ dependency than suggested by linear perturbation theory.  
In addition,
we show that both two-point functions for velocities disagree with
linear theory predictions out to $\sim 40\,h^{-1}$ Mpc separations.  A
pedagogical appendix is included developing linear theory notation
with respect to the two--point peculiar velocities functions.
\end{abstract}

\begin{keywords}
cosmology: theory -- cosmology: observation -- cosmology:
large-scale structure of the Universe -- galaxies: clusters: general
\end{keywords}

\section{Introduction}

The growth of galaxy and galaxy cluster peculiar
velocities provides information on the growth of structure in
the gravitational instability paradigm.  Over the last decade, cosmic
velocity fields were an active area of research, both in observation
and in theory (e.g., \citet{bahcall, strauss95}). 
Bulk peculiar velocities of galaxies and clusters
were modelled as tracers of the background dark matter velocity field
and were frequently used to constrain cosmological
parameters even fairly
recently \citep{silberman, bridlezehavi, feldman, juszkiewicz, shethdiaferio, peel02a}.

As measured by the number of relevant papers and conferences,
there has been a slightly falling interest in velocity work,
in part due to observational limitations.  The fundamental plane method of
measuring galaxy peculiar velocities, based on methods such as
Tully-Fisher and D$_n$-$\sigma$, is limited by relative
intrinsic errors, which grow as a percentage of distance
\citep{jacoby92}.  For a highly selected subsample, the errors can be
as low as $\sim10$ per cent, but in general the error is closer to
15-20 per cent
of the distance.  This has limited the direct use of velocities for
cosmology to redshifts of $z\simeq 0.024$, roughly a comoving
distance of $70\,h^{-1}$ Mpc (e.g., \citet{bridlezehavi}) 
(throughout this paper,
$h$ is the dimensionless hubble parameter such that today,
$H_0 =h\ 100\ \kms\,\Mpc^{-1}$).

In contrast, peculiar velocities derived from doppler-shifted
(`kinetic') Sunyaev-Zel'dovich effect (kSZ) spectra are subject to entirely
different systematic and intrinsic errors.  The kSZ effect probes
the hot gas within the cluster and represents a noisy estimate of the
cluster's bulk motion \citep{holder02}.  Complex motions of the
hot gas (cold fronts, cooling flows, \etc) further increase the noise
\citep{nagai03}.  In addition, measuring the
kSZ signal, which only reflects the motion of the innermost part of
intracluster gas, is a challenging observational
effort \citep{knoxchurch,diaferio05}; some recent experiments have
achieved limited success \citep{benson03}.  
Nevertheless, as the kSZ effect is not
redshift limited in principle (errors may grow {\em indirectly} 
with the distance as a function of cluster--parent halo evolution), it
is a very promising technique for constraining parameters.

With the view that kSZ observational difficulties can be mitigated by
sheer numbers and clever signal separation techniques, the question
remains as to whether we are applying the right theoretical velocity
models.  Typically, linearized first order perturbation theory 
(``linear theory'' in this paper) has been used
for large scale velocity fields, although
modes at the galaxy scale
($\sim$ Mpc) are non-linear and even inter-galaxy scales ($\sim$10
Mpc) are quasilinear for $z\sim 0$.  This has been justified by
only relying on the theory at larger scales where it is reasonable to 
invoke the stable clustering regime.

Yet clusters are assumed to exhibit large fluctuations at large scales
and are clearly biased samples in the linear regime.  How does this
affect the model?  In other words, what does selection
bias do to the statistics of these velocities?
In light of the growing body of work on non-linear halo evolution in
simulations using the `halo model' \citep{mowhite96, shethtormen99}, 
peculiar velocities are due for a similar
detailed examination.  This is especially true for so-called
`precision cosmology' efforts when constraining parameters such as
$\Omega_m$ and $\sigma_8$.  This might seem obvious in dealing with
galaxy peculiar velocities.  But it is also true for the streaming
motions of galaxy clusters, whose rarity implies that 
environment dependence, biasing,
and selection effects may be more important than for galaxies.

For instance,
simulated clusters (large mass dark matter halos) 
show \rms peculiar velocities
that depart from linear theory \citep{colberg00}.  An earlier paper
found that linear theory predictions were `somewhat lower' than
N-body results  and noted the disagreement between
simulated and linear theory two-point functions \citep{croft95}.
Reasonable attempts to explain the excess \rms of peculiar velocities 
using simulations have been published (\cite{shethdiaferio,
hamana03}), although not specifically as functions of cosmological
parameters.   In this paper, we examine the full two-point velocity
functions through N-body simulations while varying $\Omega_m$.  
Understanding the behaviour of these functions will be necessary for
observations to yield constraints on parameters.

In \S 2, we provide a brief pedagogical discussion of linear theory peculiar
velocities and discuss why linear theory is overly simplistic.  We
address the peak-background split approach and examine how selecting
over the peaks in the density field affects theoretical predictions.
In \S 3 we discuss our simulations.  In \S 4 we summarize our
results.  We discuss our results in \S 5.  Our conclusions in \S 6
also include a discussion of the usefulness of our approach for 
parameter forecasting in general.  An Appendix is included in this
paper which outlines how to calculate the two-point correlation
functions for velocities of peaks.

\section{Linear Theory}

\subsection{The two-point correlation tensor}

We derive the two-point velocity functions in real space (\eg,
\cite{gorski88}). (For mathematical elaboration, see also the Appendix
in this paper.)
This is crucial to properly construct the velocity
correlation matrix used in any constraint analyses, as
cross-correlations between velocities must be taken into
account.

By ``linear theory'', we mean that we begin with an initially Gaussian
distributed field and evolve it in the linear regime of the
gravitational instability paradigm via perturbation theory within a
standard Friedman-Walker-Robertson universe.  

With that simplification, the continuity equation
\be
\label{eqn:continuity}
\veck\cdot \vecv_k = i\dot{\delta}_k
\ee
implies that the curl-free $v_k$ will grow as the time derivative of
the density field, where $\delta_k$ is the comoving mode of the
density contrast $\delta\rho/\rho$ and $v_k$ is the Fourier velocity
component parallel to that mode.  The overdot is the
conformal time derivative.

The two-point statistic encompasses correlations between two
vectors, so we have the nine-element tensor:
\be
\label{eqn:tensorstart}
\Psi_{ij}(r) = \langle v_i(\vecx)v_j(\vecx + \vecr)\rangle\ \ \ \
 (i,j=1,2,3)
\ee
where we invoke isotropy and homogeneity so that $\Psi$ can only
depend on the comoving distance $r = |\vecr|$ between pairs of
velocities at {\em comoving} positions $\vecx = \vecr_1$ and $\vecx +
\vecr = \vecr_2$. The $\langle\rangle$ brackets refer to an ensemble
average.  It is
straightforward to derive the two-point radial velocity
function. On average, there are only two nontrivial correlations: one
for the components of the velocities parallel to the line between them
($\Psi_\para$), and one for the components perpendicular to that line
($\Psi_\perp$). We obtain:
\be
\label{eqn:psitheory}
\Psi(r) = \Psi_\perp\hat{{\mathbf I}} + (\Psi_\para - 
\Psi_\perp)\hat{\vecr}\hat{\vecr}
\ee
where $\hat{{\mathbf I}}$ is the identity tensor and $\hat{\vecr}$ is the
unit vector along $\vecr$.

\begin{figure}
\plotone{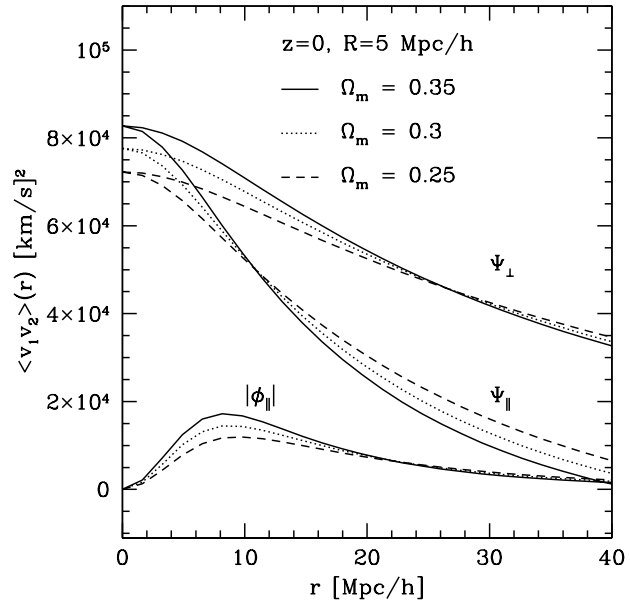}
\caption{Linear theory velocity correlations of a Gaussian distributed
  field smoothed with a $5\,h^{-1}$ Mpc window function -- all other parameters are
  held fixed for a flat $\Lambda$CDM universe. From top to bottom:
  $\Psi_\perp$, $\Psi_\para$, $\phi_\para$ ($\Phi_\para$ is discussed
  in \S 2.3); solid, dotted and dashed lines show $\Omega_m$=0.35, 0.3
  and 0.25 respectively.
  \label{fig:theoryA} } 
\end{figure}

Fig.~\ref{fig:theoryA} shows how $\Psi_\para$ and $\Psi_\perp$ depend
on comoving distance between two points for three
different flat $\Lambda$CDM cosmologies (parameters other than
$\Omega_m$ fixed; $\Phi_\para$ is discussed below in \S 2.3).  
Note the dependence on $\Omega_m$ is fairly degenerate
with normalization such as by $\sigma_8$.  In addition, these
functions have been convolved with smoothing tophat window
function to account for the velocity of an extended section of the
field, rather than for a point particle.

For a given pair of {\em radially projected} velocities separated 
by angle $\theta$ on the sky, $v_r(\hat{\gamma}_1,r_1),$
$v_r(\hat{\gamma}_2, r_2)$, the two-point correlation is:
\bea
\label{eqn:psi}
\Psi_{12} & \equiv & \langle v_{r1}v_{r2}\rangle 
 =  \hat{\gamma}_1\cdot\Psi\cdot\hat{\gamma}_2 \nonumber\\
& = &\Psi_\perp\cos\theta + (\Psi_\para - \Psi_\perp)f(\theta,r_1,r_2)
\eea
where:
\be
\cos\theta = \hat{\gamma}_1\cdot\hat{\gamma}_2
\ee
and
\be
f(\theta,r_1,r_2) = {(r_1^2 + r_2^2)\cos\theta - r_1r_2(1 +
  \cos^2\theta) \over r_1^2 + r_2^2 - 2r_1r_2\cos\theta}.
\ee
For the extreme cases of either two positions lined up along
the line of sight:
\bea
&\theta \rightarrow 0  {\rm\ and\ } r_1 \neq r_2\ \ \  f \rightarrow 1
\nonumber\\
&\Psi_{12}  = \Psi_\para
\eea
or for two positions at the same radial distance, but separated by a
small angle:
\bea
&\theta \neq 0 {\rm\ but\ small\ }  {\rm and\ } r_1 = r_2\ \ \  f
\rightarrow -{\theta^2\over 4}\nonumber\\
&\Psi_{12} = \left(1 - {\theta^2\over 4}\right)\Psi_\perp -
            {\theta^2\over 4}\Psi_\para \approx \Psi_\perp.
\eea

The zero-lag for either two-point function is related to the \rms
velocity:
\be
\label{eqn:sigma}
\sigma_v^2 = \langle v^2\rangle =
\langle\vecv(\vecx)\cdot\vecv(\vecx)\rangle = \Sigma_i \langle
v_i^2\rangle = 3\Psi_{(\para\ or\perp)}(0)
\ee
where we note that the literature often uses $\sigma_v^2$ and $\langle
v^2\rangle$ interchangeably.

\subsection{Arguments against linear theory}

The na\"ive idea that linear theory in the field will predict the
velocities of galaxies should be suspect, though this has been the
norm in the past (with some notable exceptions, \eg, \citet{ma01},
\citet{silberman}).
Galaxies are not only non-linear objects themselves, their velocities
are often responding strongly in the non-linear regime since perhaps as
many as $\sim 20$ per cent of them are in bound groups.  Furthermore, they most
likely represent biased objects compared to the general behaviour of
the dark matter background.  
Data from large scale surveys has been consistent with modelling this
bias as approximately linear for $L_*$ (and dimmer) galaxies
\citep{seljak04a}, though simulations suggest the bias is expected to increase
for larger sized haloes \citep{seljak04b}.

Clusters are on average inherently
rare objects as modelled by any Press-Schecter
type formalism.  Although a $10^{14}\,h^{-1}$ M$_\odot$ cluster began in a
comoving volume of radius $R\simeq 7\,h^{-1}$ Mpc ($\Lambda$CDM), few
such volumes have clusters.  In fact, a cluster is found at late
times (on average) in a radius $\sim$4 times that size, \ie, a volume
64 times the original source volume for the cluster.  This implies
that galaxy clusters as a selected sample
should have a bulk motion which is responding to long wavelength modes
which have not undergone collapse and are therefore well-modelled in
the linear approximation.

The statistical rarity of high mass haloes is presumed to be a source
of a suppression factor to their \rms velocity as predicted by
the excursion hierarchy approach of \cite{BBKS}.  Modeling
clusters  as originating from the 3-$\sigma$ (or greater) end of the
density peak distribution, the peak \rms velocity is given in
linear theory by:
\be
\label{eqn:peakstat}
\sigma_p^2(R) = \left(1-{\sigma_0^4(R)\over\sigma_1^2(R)
\sigma_{-1}^2(R)}\right)\sigma_v^2(R)
\ee
where:
\bea
\label{eqn:sigman}
\sigma_n^2(R,\eta) &=& \int {k^2dk\over 2\pi^2} P(k,t) k^{2n}W^2(kR)\\
&=& D^2(\eta)\int {k^2dk\over 2\pi^2}|\delta_{k,0}|^2 k^{2n}W^2(kR).\nonumber
\eea
Implicit in the final form above is the linear theory assumption that
$\delta(\vecx,\eta)$ is separable; thus
$\delta(\vecx,\eta)=D(\eta)\delta_0(\vecx)$ and we choose the growth
function $D$ to be normalized to one today. ($D=a$ for an
$\Omega_m$=1 cosmology).  This assumption is reasonable if we smooth
on large enough scales to ignore nonlinear processing on small scales,
which is the purpose of the window function $W(kR)$.  The window
function is usually chosen to
be the Fourier transform of either a Gaussian or tophat envelope.
When the time ($t$ or $\eta$) 
is not specified, $D$, or $P(k)$ assume values at
redshift zero, \ie, today.  For examples of this notation, the
cosmological parameter $\sigma_8 = \sigma_0(8\,h^{-1}$ Mpc), and the
$\sigma_v$ of Eqs.~(\ref{eqn:sigma}) and (\ref{eqn:peakstat}) above is
$\propto \sigma_{-1}$.

The \rms velocity of rare, massive objects is apparently suppressed by a few
to ten percent compared to background as implied by Eq.~(\ref{eqn:peakstat}).  
\citet{colberg00}
show that although this statement agrees for the velocities of the
density peaks at early times, large mass haloes evolved to low
redshift which formed around these peaks actually have a higher
\rms velocity than linear theory would have predict by as much
as 40 per cent.  This discrepancy is observed in our simulations even at
moderate redshifts ($z\gtrsim 0.6$) in cluster formation and will be
discussed in \S 5 below.  See the Appendix for more
detailed calculations to predict peak-peak velocity correlations as a
natural extension to Eq.~(\ref{eqn:peakstat}); however, it is fairly
clear from simulations that the peak-background split approach is not
accurate at late times, as evidenced for example with the
\rms peak velocity above.

How does one take into account the bias for dense objects, like
galaxies and clusters, caused by their being in an overdense region?
In general, an object's peculiar velocity is greatly affected by its
environment: haloes in overdense regions typically move faster than
those in less dense ones \citep{colberg00,shethdiaferio,hamana03}.
Although the average intracluster distance is large, the likelihood of
finding a cluster near another is high, and of finding a cluster near
a large overdensity of galaxies and groups is very high.
If environment plays an
important role for the evolution of galaxy peculiar velocities, it
must therefore play an even more important one for clusters.

\subsection{Momentum correlations}

A heuristic way to see the effect of this selection bias is to re-examine
Eq.~(\ref{eqn:continuity}) within the  context of 
linear theory.  The full form, after separating out the background
solution and regardless of the amplitude of $\delta(x)$ is:  
\be
\label{eqn:exact}
\dot{\delta} + \nabla\cdot[(1+\delta)\vecv] = 0\ .
\ee
This suggests we might consider the statistic:
\bea
\label{eqn:momentum}
\tilde{\Psi} & = & 
\langle\tilde{\vecv}(\vecx)\tilde{\vecv}(\vecx+\vecr)\rangle\nonumber\\
& = & \langle \vecv(\vecx)(1+\delta(\vecx)) 
\vecv(\vecx+\vecr)(1+\delta(\vecx+\vecr)\rangle \\
& = & \langle \vecv_1\vecv_2\rangle + \langle \vecv_1\vecv_2\delta_1
\rangle + \langle \vecv_1\vecv_2\delta_2 \rangle
+ \langle \vecv_1\delta_1 \vecv_2\delta_2\rangle\nonumber
\eea
where the subscripts `1' and `2' correspond to the arguments
$\vecx$ and $\vecx+\vecr$ respectively.  This is a `momentum
correlation', \ie, weighting the velocity by the density in the
region.

Evaluating Eq.~(\ref{eqn:momentum}) using arguments within 
linear theory solely to build intuition (\ie, three-point functions
are zero, ignoring the connected part of the four-point function,
\etc) leads to:
\be
\label{eqn:momentum2}
\tilde{\Psi}(r) = \Psi_\perp(1+\xi(r))\hat{{\mathbf I}} + ((\Psi_\para -
\Psi_\perp)(1+\xi(r))+\Phi_\para)\hat{\vecr}\hat{\vecr}
\ee
with $\Psi_\perp$ and $\Psi_\para$ as before, and:
\bea
\label{eqn:phi}
\Phi_\para & = & \langle\vecv_1\delta_2\rangle\langle 
\vecv_2\delta_1\rangle\nonumber\\
& = & -{\dot{D}_1\dot{D}_2D_1D_2\over 4\pi^4}\left\{\int
dk\ |\delta_{k,0}|^2 k j_1(kr) W^2(kR)\right\}^2\ .
\eea
$\xi(r)$ is the usual Fourier transform of the power spectrum:
\be
\xi(r) = \langle \delta_1\delta_2\rangle = D_1D_2 \int 
{k^2dk\over 2\pi^2}|\delta_{k,0}|^2j_0(kr)W^2(kR).
\ee
Fig.~\ref{fig:theoryA} shows the behaviour of $|\Phi|$ at $z$=0.

So how does Eq.~(\ref{eqn:momentum2}) compare to the unweighted
model, Eq.~(\ref{eqn:psitheory})?  The extra factor of $\xi$
boosts the correlations for velocities at short ($\sim 10\,h^{-1}$ Mpc)
separations.  The new factor of $\Phi_\para$ which is zero at both zero
lag and large separations boosts the anticorrelations in $\Psi_\para$
for a characteristic separation.  This is not presumed to be exact, 
but will aid discussion of results in \S 5.

By definition, linear theory becomes invalid when $\delta\sim
1$ which forces the equations in Fourier space to mix modes.  
This is where numerical simulations become essential.  

\subsection{Previous efforts regarding the velocity \rms}

Extensive work by \cite{shethdiaferio} and \cite{hamana03} comparing
halo velocities in simulations to linear theory showed that the local
environment of a halo had a heavy influence on the evolution of its
velocity.  Specifically, the bias predicted by the halo model
suggested that halo velocities would likely be boosted as if they had
evolved in a higher-$\Omega_m$ universe (see Fig.~\ref{fig:theoryA})
because haloes are typically found in overdense regions.

In one approach \citep{shethdiaferio}, it was suggested that a typical
halo speed today would be related to the linear growth velocity
(evolved from redshift 20 until today) boosted by the local density in
the region: 
\be
v_0 = (1+\delta)^{\mu(R)}{\dot{D}(\eta_{20})\over \dot{D}(\eta_0)} v_{20}
\ee
where $\delta$ is smoothed over a region of radius $R$ using a
Gaussian window function. They found that $\mu$ was naturally tied to
the choice in smoothing radius, and for their simulations, fit:
\be
\mu(R) = 0.6{\sigma^2(R)\over \sigma^2(10\,h^{-1}\,\Mpc)}.
\ee

Following this guideline, the second approach \citep{hamana03} found a
similar result phrased as the \rms velocity (essentially
the same statistic: see the final paragraph in \S 2.1 above):
\be
\sigma^2_{halo}(M,\delta) = [1 + \delta(R_{local})]^{2\mu(R_{local})}
\sigma_v^2(M)
\ee
although neither group found the velocities to have a strong
dependence on halo mass.

In the first work, the choice of $10\,h^{-1}$ Mpc  as a
reference scale for smoothing is motivated by the fact that it
roughly represents the transition from linear to non-linear regimes
today as measured by $\sigma(R)\sim 1$.
According to the second work, deciding how to choose $R_{local}$ is
primarily an {\em ansatz}.

The velocity statistics in previous papers investigated 
a wider range of dark matter halo masses.  We
will focus on the statistics of only the largest mass haloes by
imposing a mass cutoff suggested from models of the
Sunyaev--Zel'dovich effect \citep{carlstrom02}.  In addition, we will
examine this selection effect on the full two-point functions, rather
than simply the zero-lag.

\section{Simulations}

\subsection{ART code}

We used an Adaptive Refinement Tree (ART) N-body code \citep{kravtsov99}
which,
like its predecessor Particle Mesh (PM) N-body codes,
integrates trajectories of
collisionless particles by solving the Poisson equation. Unlike PM
codes, ART allows for a 
hierarchy of refinement meshes where collapsed objects require more
resolution. 

ART employs standard
particle-mesh techniques to compute acceleration grids in order to
advance
particle coordinates and velocities in time.  A regular cubic grid covers
the entire computational volume and defines the initial minimum resolution of
the simulation.  This grid is then refined where the density contrast
is higher to form higher resolution sub-meshes in those regions of 
interest.  The main computational loop of the integration
consists of: {\bf (1)} density assignment for all existing meshes; {\bf
(2)} running the gravitational solver; {\bf (3)} routine updating particle
positions and velocities; {\bf (4)} modifications to the mesh
hierarchy.

\subsection{Halo finder algorithms}

The ART codes we used produce files of particle positions and velocities which 
were subsequently analysed for the presence of haloes.  
The basic problem of halo finding in a simulation is that there are 
no clear boundaries for haloes.  There is no single perfect
algorithmic definition of a group or mass of a group.  

Many halo finding
algorithms exist, but tend to fall into two categories:
the friends of friends type (FoF) linked-list type approaches where 
particles are identified with a halo if they are within a certain 
chosen distance of each other
\citep{efstathiou85}; and overdensity methods such as DENMAX which calculates
the density as a function of a grid and identifies to which local maximum
each particle belongs \citep{bertschinger91}.  We used a relatively 
recent method 
named HOP \citep{eisenstein98} which
follows the logic behind overdensity methods yet includes `hopping' to
nearest neighbors {\em \`a la} FoF methods.  Instead of calculating the 
density on a grid, a density is associated with each particle.  Then a
search is conducted
for the highest density nearest neighbor until a particle is its own 
densest neighbor.  All particles which trace to the same such particle are
grouped.  A followup `regrouping' then reunites any sufficiently bound
haloes which happen to contain two (or more) 
local maxima such that the initial hopping misidentified them as separate
haloes.

\subsection{Virial radius}

After HOP was used to find the haloes, a crude spherical overdensity 
method was applied to restrict the statistics to different cutoff
radii.  Real measurements of cluster peculiar velocities via the
kSZ effect will be restricted to the baryons at the core but
are likely to represent at best the bulk motion of particles
`trapped' within the virial radius \citep{holder01}. We define the
virial radius by beginning nearest the central overdensity of a halo
and including particles at every increasing radii until the
overdensity within that radius is 180 times the background density.

\subsection{Preliminaries}

The first questions to answer using N-body simulations
were to determine: {\bf (1)} how many high-mass haloes (presumably
hosting clusters) were needed;
{\bf (2)} how big the simulated volume should be; and {\bf (3)} how much mass
resolution was required for each halo.   This phase was completed
using approximately  10,000 hours of processor time on the
the IBM SP computer, `Seaborg', at the National Energy Resource
Computing Facility at Lawrence  Berkeley National Laboratory.

\subsubsection{Number of high mass haloes (clusters)}

We relied on linear theory predictions as a rough guide in determining
the number of high mass ($M\gtrsim 3\times 10^{14}$ M$_{\odot}$) haloes
we would need to achieve an 
error variance on the order of a percent for $\Omega_m$.  

For effectively uncorrelated cluster velocities, from, \eg,
a very sparse survey, we estimate the expected error variance
on $\Omega_m$.  From a measurement of $N$ clusters with their 
peculiar velocity variance represented by the zero-lag value of
either two-point function ($\Psi_0(z_i))$:
\bea
(\Delta \Omega_m)^2 &=&  \left(\sum_i \left({\partial \Psi_0(z_i) \over
\partial \Omega_m}
\right)^2 {1 \over 2(\Psi_0(z_i)+\sigma_{v,noise}^2)^2 }\right)^{-1}\nonumber\\
&\simeq& {800\over N}(.01)^2
\eea
where the last equality assumes $N$ clusters with $\sigma_{v,noise}^2\ll \Psi_0$
all at $z$=1, and $\partial \ln \Psi_0 /\partial \Omega_m \simeq 5$ 
\citep{peel02a}.  Thus, on the order of 1000 clusters would be apparently sufficient to
constrain $\Omega_m$ to a few percent.  Current and future
cluster surveys expect to detect on the order of 10,000 clusters 
through the Sunyaev-Zel'dovich effect \citep{carlstrom02}.

\subsubsection{Volume}

To determine the necessary volume to find these haloes, 
we followed Jenkins \citep{jenkins01} fitting formula for the `universal 
mass function':
\be
f(M) = 0.315\exp(-|\ln\sigma^{-1}+ 0.61|^{3.8})
\ee
where $\sigma^2(R)$ is the usual smoothing of the power spectrum
with a window function $W(kR)$, and the number of haloes of mass $M$ 
at a redshift $z$ is:
\be
\label{eqn:dndm}
{dn\over dM}(M,z)dM = 2{\bar{\rho}\over M}\nu f(\nu(M))d\nu
\ee
where
\be
\nu = \delta_c/\sigma(M)
\ee
and $\delta_c$ is the critical value of
a spherical overdensity at turnaround time.  The universality
referred to is due to the functional form of $f$ and is not as useful for
our purposes as Eq.~(\ref{eqn:dndm}) above.  
For 300 high mass clusters ($3\times 10^{14}\,h^{-1}$ M$_\odot$)
at a redshift of $z\sim 0.6$,
we required a fairly large volume of ($850\,h^{-1}$ Mpc)$^3$.  The steepness
of the halo mass function would then guarantee $\sim 10^3$ clusters with
mass greater than $2\times 10^{14}\, h^{-1}$ M$_\odot$ at $z\sim 0.6$.

\subsubsection{Number of particles per halo}

We simulated a volume with the same initial
conditions but with three different mass resolutions to see how many 
particles were needed to resolve halo velocities.
For this convergence test, we used smaller boxes of $150\,h^{-1}$ Mpc per side
for speed.  At this size, we expected very few haloes above
$10^{14}\,h^{-1}$ M$_\odot$ at a redshift of $z \sim 0.6$.

Beginning with $256^3$ cells, we used number of particles $64^3$, $128^3$
and $256^3$ and tracked the velocities of the top five haloes as they
became more resolved, as well as the \rms of the entire population of
haloes.  From this convergence test, it became clear that the number of 
dark matter particles 
required to resolve a velocity was approximately 70, which meant that
for an $850\,h^{-1}$ Mpc sized box, $(256)^3$ particles would be
sufficient to resolve the velocities of the 
largest haloes, \ie, those haloes most likely to contain massive clusters.  
Eight $425\,h^{-1}$ Mpc per side simulations would cover the same volume and run much
faster, but with the loss of the $k=2\pi/850\,h$ Mpc$^{-1}$ mode.
We ran multiple $425\,h^{-1}$ Mpc per side
simulations with $128^3$ particles and compared results with one $850\,h^{-1}$
Mpc using $256^3$ particles.  We found that 
losing the low-$k$ mode ($k=2\pi/850\,h$ Mpc$^{-1}$) had a negligible effect
on velocity statistics.  

For our chosen resolution of $128^3$ particles realized in ten ($425\,h^{-1}$
Mpc$)^3$ volume boxes, $m_p = 3.0\times 10^{12}\,h^{-1}$ M$_\odot$
($\Lambda$CDM, $\Omega_m$=0.3).  
This means that any halo identified with approximately 60
particles ($1.8\times 10^{14}\,h^{-1}$ M$_\odot$)
is a halo capable of hosting a cluster.  The simulations were run
using three values of $\Omega_m$ ($0.25,0.3,0.35$) in flat $\Lambda$CDM
cosomologies with all other parameters fixed
on the  UK National Cosmology Supercomputer in Cambridge.  Each
realization took less than one week on eight Altix 3700 Itanium2
processors.  (Eight Altix processors was approximately ideal for this
number of particles and timesteps.)  We discuss the broad usefulness
of this approach in \S 6.

\section{Results}

\begin{table}
\caption{Number and Peculiar Velocities of High Mass Haloes
from N-body Simulations}
\begin{tabular}{@{}|c||c|c|c|c|c|c|c|@{}}
\hline
 & \multicolumn{3}{|c||}{$N(>10^{14.26}\,h^{-1}$ M$_\odot)$} & &
 \multicolumn{3}{|c|}{$\sigma_v$ $[\kms]$}\\
$z$ & \multicolumn{3}{|c||}{$\Omega_m$} & & \multicolumn{3}{|c|}{$\Omega_m$}\\
\cline{2-4} \cline{6-8}
 & 0.25 & 0.3 & 0.35 & & 0.25 & 0.3 & 0.35 \\
\hline
\hline
0.0   & 9344 & 12618 & 16541 &  & 532 & 539 & 534 \\
0.25  & 6211 & 8235  & 10530 &  & 554 & 552 & 543 \\
0.667 & 2320 & 3009  & 3790  &  & 555 & 542 & 536 \\
1.5   & 109  & 131   & 146   &  & 563 & 522 & 462 \\
\hline
\end{tabular}
\end{table}

Table 1
summarizes the the \rms peculiar velocity and number of haloes found
above the cutoff mass as a function of cosmology in our total volume (10
$\times(425)^3 \simeq$ ($915\,h^{-1}$ Mpc$)^3)$. For brevity,
we will initially display our results in the context of only one
cosmology, a flat $\Lambda$CDM with $\Omega_m$=0.3. We first consider
the Gaussianity of the one-dimensional velocity distribution (and the
related Maxwellian speed distribution) to justify error estimates.
Then we show the zero lag portion of the two-point functions: the
\rms peculiar velocities as a function of mass and density.  
Following this, we reveal the
primary result of this work: the redshift  evolution of the two-point
functions for haloes above a cutoff mass. At that point, we will also
introduce our results from two other cosmologies, flat $\Lambda$CDM with
$\Omega_m$=0.25 and 0.35.  We conclude this section by showing the
bulk velocity history of the particles which make up the zero
redshift haloes above a cutoff mass.  We discuss and 
explain the results in \S 5.

Only haloes with a minimum mass of $1.8\times 10^{14}\,h^{-1}$
M$_\odot$ (60 particles in our fiducial $\Omega_m$=0.3 runs) were used
to ensure reliable velocities (see \S 3.4.3 above).  Our conclusions 
will not depend on this low mass estimate for a cluster capable of
producing a measurable Sunyaev--Zel'dovich effect; in fact, they would
be more dramatic for higher mass cutoffs.  After identifying
haloes by their member particles as described above, we calculated the
bulk flow peculiar velocity by averaging particle velocities within
the virial radius, defined as the spherical radius for which the halo
is 180 times the background density.  The resulting velocity vector is
then used to calculate all halo velocity statistics.

To understand how appropriate linear theory is for certain scales (and
as a consistency check on our simulations), we also track the average
velocities and two-point functions for the field.  To do this, each
simulation is partitioned into equal-sized boxes and the particle
velocities within each box are averaged to create a `bulk' peculiar
velocity.  The process is repeated for different smoothing lengths.
We coin these partitions `miniboxes.'  The statistics of these
minibox bulk velocities represented a simulated version of linear
theory, convolved with cubic window functions.

This partitioning was repeated for different nominal linear sizes of
$L = 8$--$128\,h^{-1}$ Mpc by powers of two.  Since the simulations were of
linear size $425\,h^{-1}$ Mpc, these lengths were rescaled due to roundoff
to  8.02, 16.3, 32.7, 70.8, and $142\,h^{-1}$ Mpc so that the entire
volume would be sampled.  These values correspond
to tophat-sphere smoothing volumes with radii of 4.98,
10.1, 20.3, 43.9 and 87.9 $h^{-1}$ Mpc.  This was tested by examining the
theory calculation with either a boxcar window function or a tophat
window function.  In either case the same theoretical results were
obtained when the volumes of the respective cubes or spheres in real
space were equivelant.  Each minibox was then assigned a density based
on the number of particles found inside: $\delta_i = n_i/\bar{n} -
1$.  Attempts to look at miniboxes of  smaller extent were limited by
the spatial resolution of the simulation.

Three things should be mentioned about how the linear theory was
calculated.  First, the apparent dependence on mass shown in the downward
curvature of the linear theory is due to convolving the integral in
Fourier space with a tophat window function $W(kR)$, with a mass
associated with the comoving volume $4\pi R^3\bar{\rho}/3= M$ in
contrast to the smaller smoothing radii used by \citet{croft95}.
Secondly, the suppression factor mentioned in \S 2.2 above is
completely ignored as we are comparing linear theory only to the
entire particle field through the minibox statistics.
Finally, the integration limits in Fourier space were
chosen to match the simulation: the lowest frequency
associated with the size of the box represents the lower bound on the
$k$-integrals and the upper limit value was chosen to reflect the Nyquist
frequency.  Initially, the Nyquist frequency was simply
2$\pi/425\times 128/2\ h$ Mpc$^{-1}\approx 1\ h$ Mpc$^{-1}$. 
While the dynamic mesh
was resolved by as many as six times at late redshifts, those new distances
represent collapsed regions and do not force much of an increase in
the Nyquist frequency for integrating the linear power spectrum.
Since we integrate with a window function, the effect of using $1\
h$ Mpc$^{-1}\ vs.\ (2^6)=64\ h$ Mpc$^{-1}$ for the upper limit is negligible.

\subsection{Gaussianity of velocity distributions}
\begin{figure}
\plotone{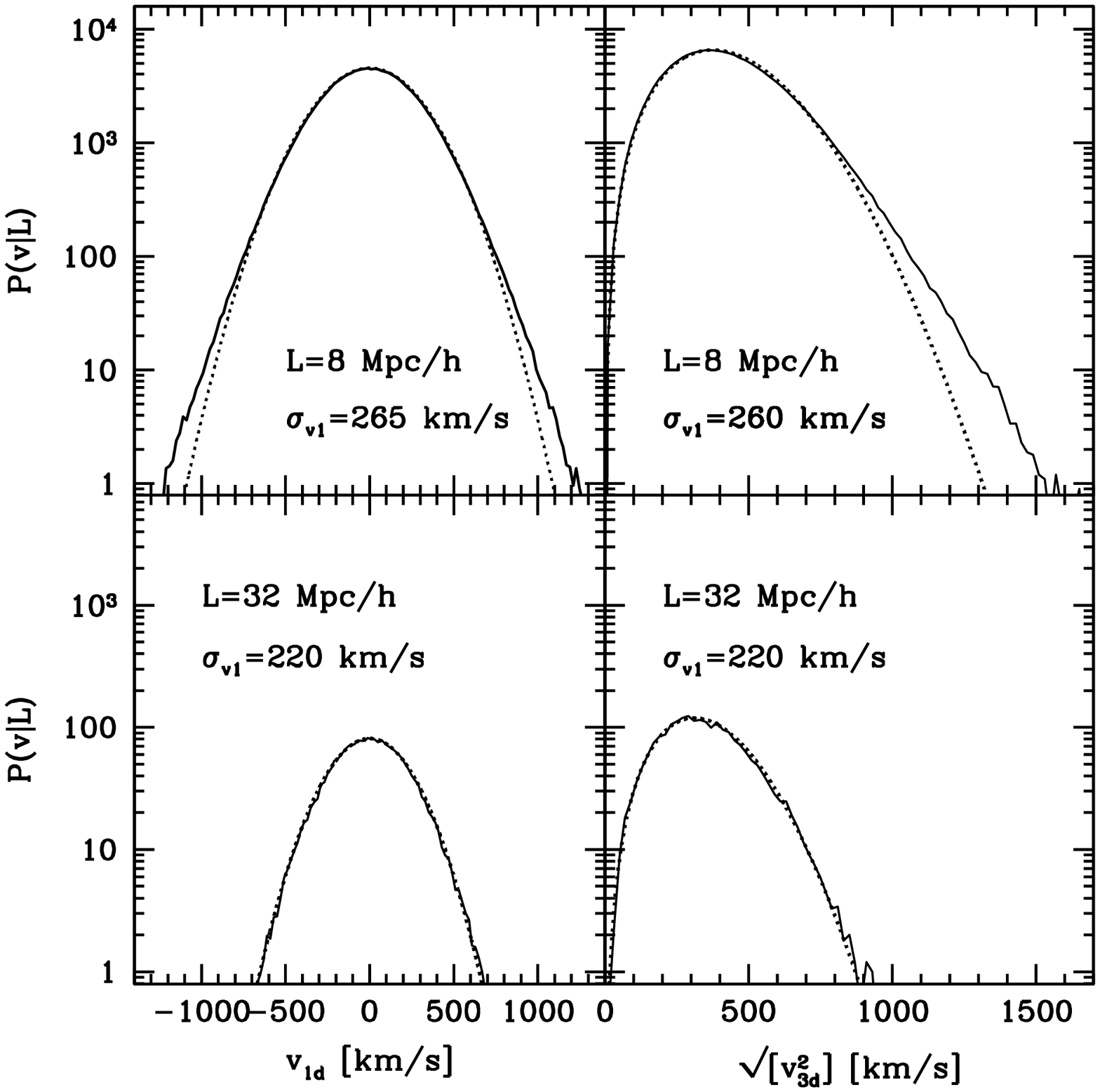}
\caption{One-dimensional velocity distributions (left)
([$v_x, v_y, v_z$]) and speed distribution (right) for miniboxes smoothed
  over $8.02\,h^{-1}$ Mpc (top) and $32.7\,h^{-1}$ Mpc at redshift 0.  Matching
  Gaussian and Maxwellian distributions for the one-dimensional
  velocities and speed respectively are shown as dotted lines.
  Recall that $v_{rms} = \sqrt{3}\sigma_{v1}$.
 \label{fig:box_1d_0.0} }
\plotone{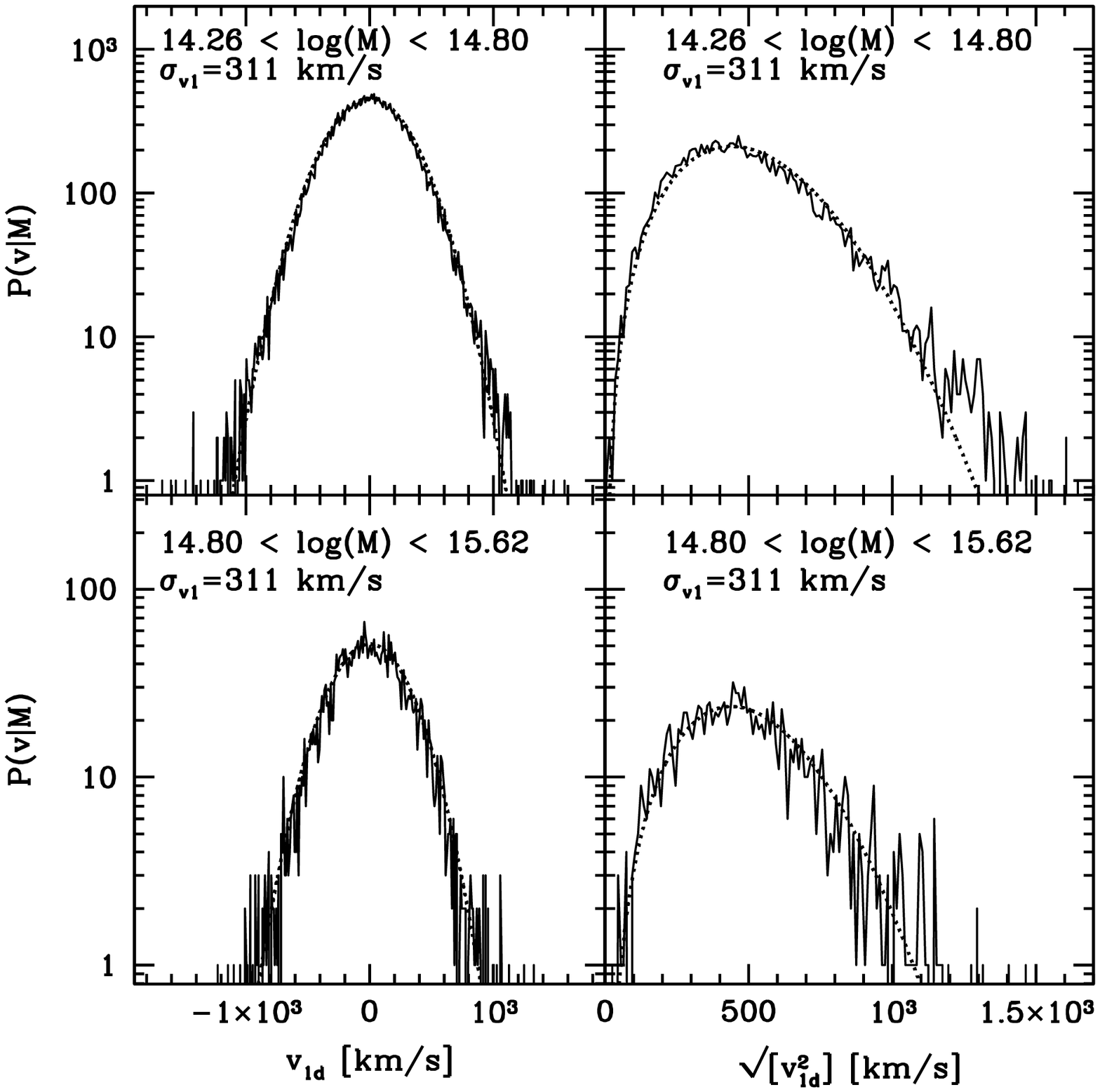}
\caption{As above, but for haloes binned in two mass ranges at redshift 0.  
 \label{fig:halo_1d_0.0} }
\end{figure}

We examine velocity distributions at redshifts $z=0$ for haloes and
miniboxes in order to understand relevant confidence intervals for
$v^2$. If the set of one dimensional velocities
$\{\vecv_i\cdot\hat{x},\vecv_i\cdot\hat{y},\vecv_i\cdot\hat{\vecz}\}$, 
is Gaussian, then $\{v^2_i\}$ should fit a Maxwellian
distribution. Figs.~\ref{fig:box_1d_0.0} and \ref{fig:halo_1d_0.0} 
below show that Gaussian
(and related Maxwellian) distributions work reasonably well for both
field (as represented by miniboxes) and halo one-dimensional velocities 
(and related 3-d speeds), even for different mass ranges.  The bias
seen in \cite{shethdiaferio} between smaller and larger
haloes is much less apparent for masses above $\sim 10^{14}\,h^{-1}$
M$_\odot$.  Both distributions are fit by a Maxwellian with one-dimensional
velocity dispersion of $311\ \kms$, \ie, an \rms of
$\sqrt{\langle{v^2}\rangle} = \sqrt{3}\sigma_{v1} = 539\ \kms$.
The shot noise from rarity at higher redshift for a large
Sunyaev--Zel'dovich survey clearly implies a transition to a Poisson
distribution when the data is binned in redshift.  This is also true
from our work, but is not shown here for brevity.

There is a high velocity tail, as predicted by \cite{shethdiaferio}.
However, for our large number of cluster-sized haloes, 
the majority of the distribution is still well enough
fitted to a Maxwellian to justify the 1-$\sigma$ errors in the figure
below given by:
\be
\langle v^4 \rangle = \left[2\over N\right]\langle v^2 \rangle^2\ .
\ee
In the case of the two-point function, this becomes:
\be
\langle \Psi^2_{\perp,\para}(r)\rangle = {2\over
  N}|\Psi_{\perp,\para}(r)\Psi_{\perp,\para}(0)|\ .
\ee

\subsection{Velocity rms}

\begin{figure}
\plotone{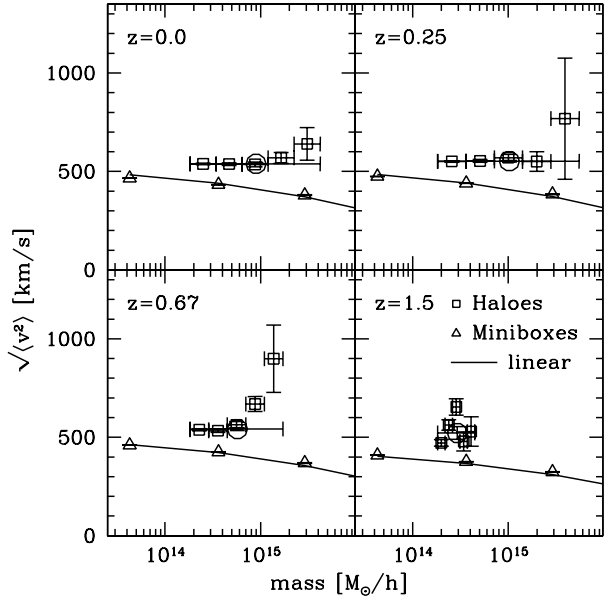}
\caption{Peculiar velocity \rms as a function of mass: boxes show halo 
  statistics; triangles show minibox statistics; 
  solid lines show linear theory.  The circle shows the average halo
  statistic for the entire mass range of interest with insignificant 
  statistical 1-sigma error bar.
  \label{fig:sigmav_vs_M} }
\end{figure}

Fig.~\ref{fig:sigmav_vs_M} shows the \rms velocities of
miniboxes and haloes {\em vs.~}linear theory predictions as functions
of mass at different redshifts.  In the mass range of interest
($10^{14}$ -- few$\times 10^{15}$) linear theory and the minibox
velocity statistics are in excellent agreement. Although not shown for
brevity, the minibox two-point functions were also in agreement with
linear theory. 

This figure also shows the discrepancy
between \rms velocities of linear theory and haloes.  This is in
rough agreement with the result that linear theory
underpredicts the \rms of cluster peculiar velocities as
represented by massive dark matter haloes by a large percent.  
For example, averaged across the mass range, the predicted value
of halo \rms peculiar velocity is 539$\pm 3\ \kms$ at redshift zero,
whereas linear theory predicts an average of 415 $\kms$ in that range,
a discrepancy of 30 per cent. In agreement
with \citet{shethdiaferio} and \citet{hamana03}, 
we find essentially no mass
dependence for the \rms peculiar velocities (within 1-$\sigma$).
Since linear theory {\em does} predict a
weak dependence on mass (larger objects should be {\em slower} as
shown by the convex line in Fig.~\ref{fig:sigmav_vs_M}) the
discrepancy found by \cite{colberg00} actually worsens for the largest
mass haloes, although clearly this is of less statistical significance.

\subsection{Two-point functions}

\begin{figure}
\plotone{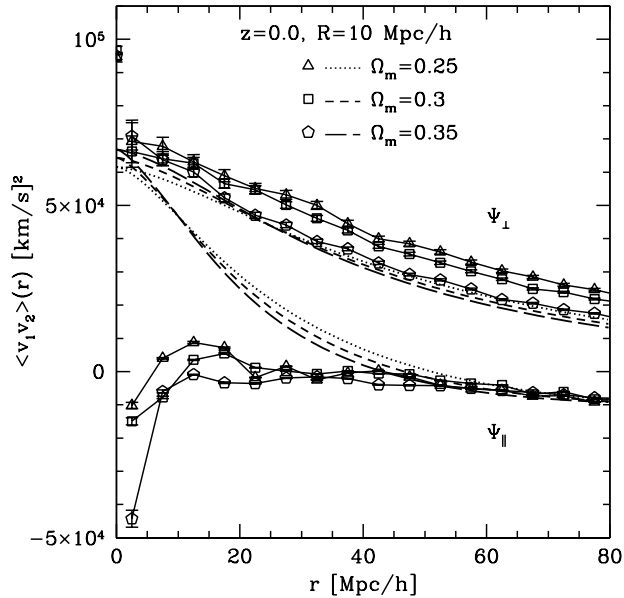}
\caption{Halo velocity two-point correlations (perpendicular and parallel
  components) at redshift $z=0.0$ for haloes above the mass cutoff
  (see text).  The dotted, short-dashed and long-dashed lines show
  linear theory and the triangles, squares and pentagons show
  simulated haloes for $\Omega_m$ = 0.25, 0.3, and 0.35 respectively.
  Note the high value for simulations at the zero lag (where
  $\Psi_\perp$ must equal $\Psi_\para$).
 \label{fig:corr_0.0} }
\end{figure}

\begin{figure}
\plotone{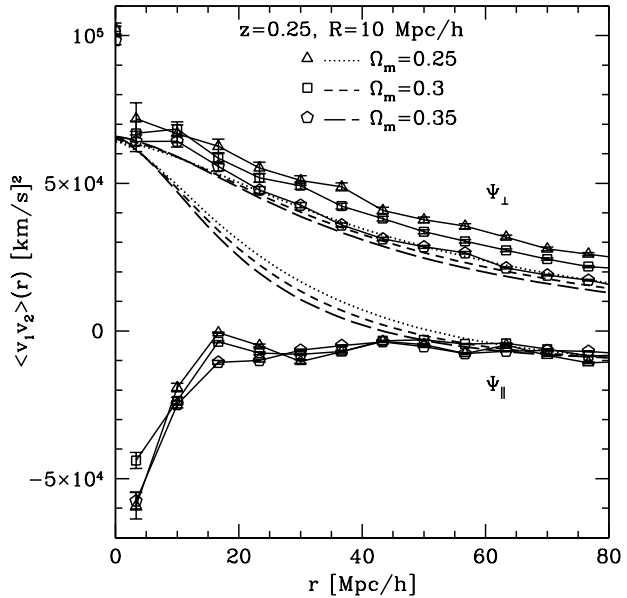}
\caption{as above, but for $z=0.25$.
 \label{fig:corr_0.25} }
\end{figure}

\begin{figure}
\plotone{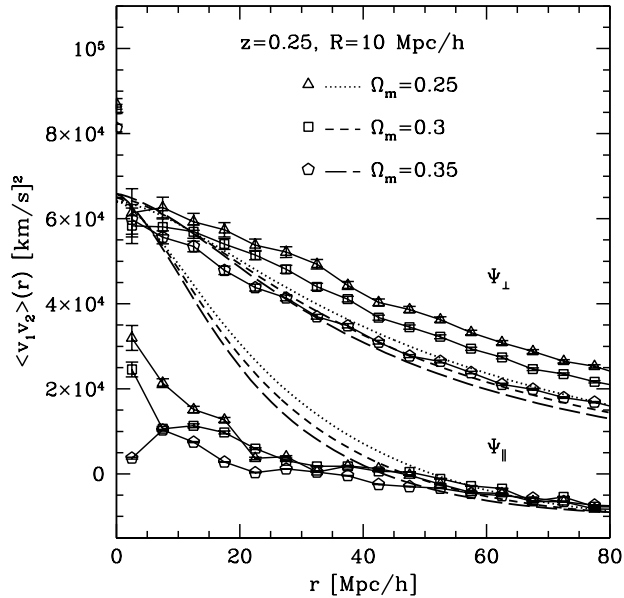}
\caption{Velocity two-point correlations at $z=0.25$
  (perpendicular and parallel components) for the groups of particles
  which will be in the haloes shown in the Fig.~\ref{fig:corr_0.0}.
  The dotted, short-dashed and long-dashed lines show
  linear theory and the triangles, squares and pentagons show
  simulated haloes for $\Omega_m$ = 0.25, 0.3, and 0.35 respectively.
  Note the high value for simulations at the zero lag (where
  $\Psi_\perp$ must equal $\Psi_\para$).
 \label{fig:hist_0.25} }
\end{figure}

\begin{figure}
\plotone{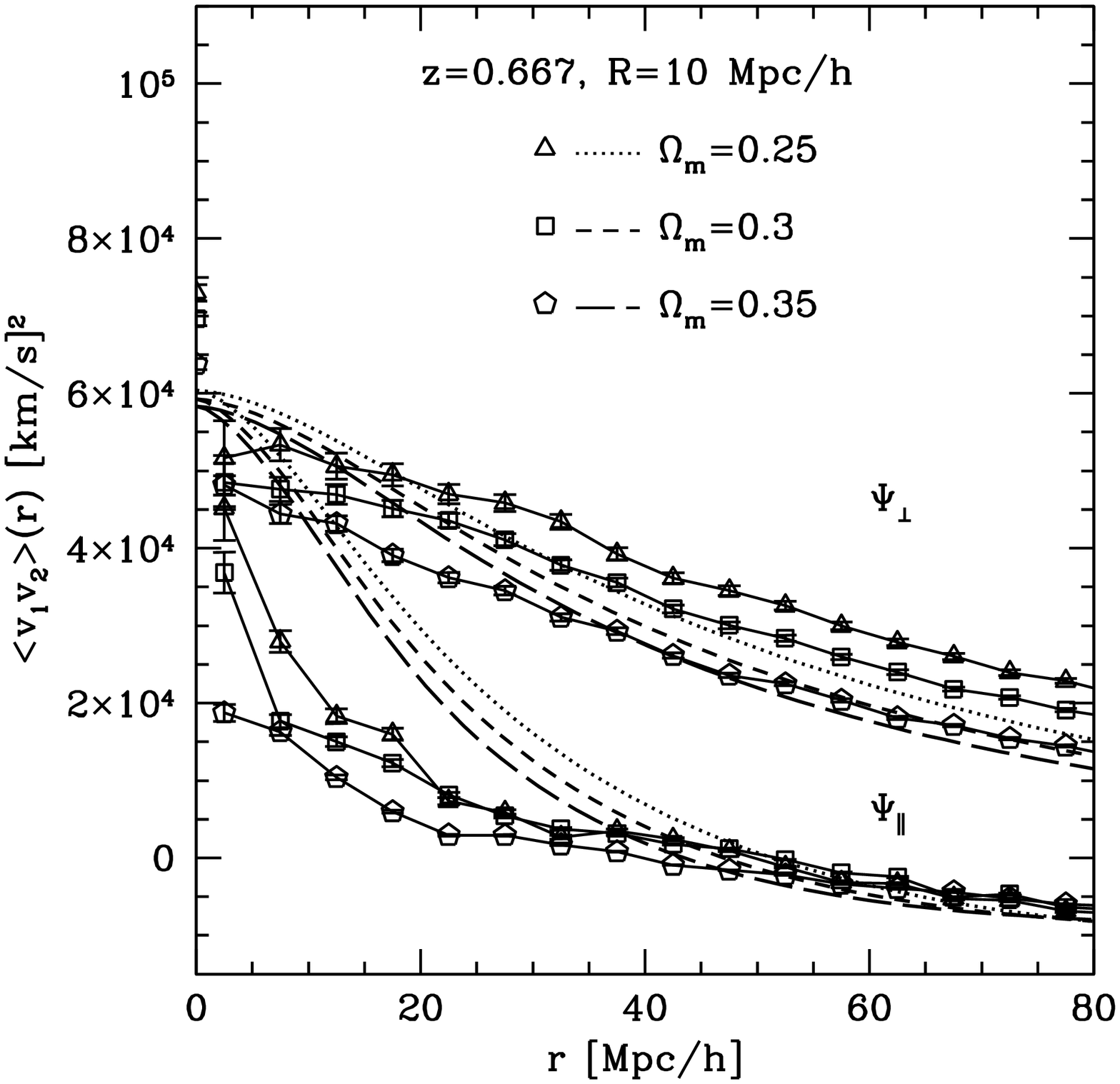}
\caption{as above, but for $z=0.667$.
 \label{fig:hist_0.667} }
\end{figure}

\begin{figure}
\plotone{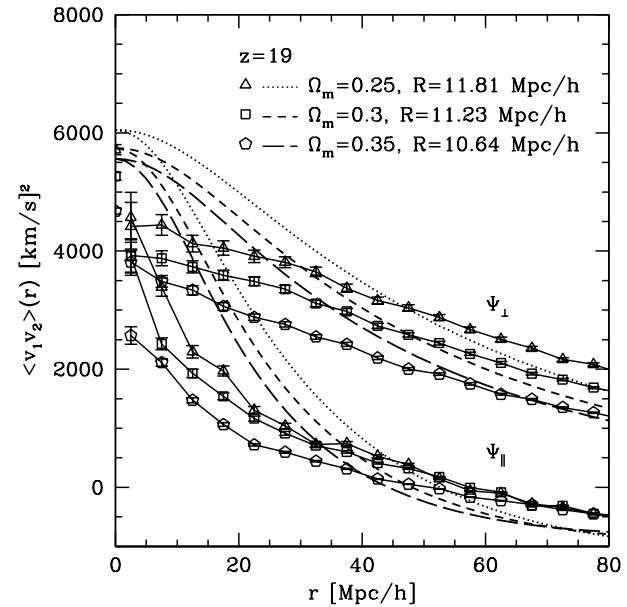}
\caption{as above, but for $z=19$.
 \label{fig:hist_19} }
\end{figure}

We show the two-point function redshift dependence it two ways.
First, motivated by obervations, we examine the haloes above a cutoff
mass which have collapsed at redshifts $z$=0.0, 0.25 and 0.667, shown
in Figs.~\ref{fig:corr_0.0}, \ref{fig:corr_0.25},
\ref{fig:corr_0.667}, respectively.  We also examine the velocity {\em
history} of the particles which will be within the virial radius of
$z$=0 haloes above the mass cutoff.  We do this by calculating particles'
`bulk' velocity as if they had formed haloes already.  We only use
those particles which will be within the virial radius at $z$=0.
Those results at redshifts $z$=0.25, 0.667 and 19 are shown in 
Figs.~\ref{fig:hist_0.25}, \ref{fig:hist_0.667}, \ref{fig:hist_19},
respectively.

For Figs.~\ref{fig:corr_0.0} through \ref{fig:hist_0.667},
we use linear theory smoothed over $R=10\,h^{-1}$ Mpc as a benchmark
for comparison, though we do not {\em a priori} expect halo velocity
statistics to be in agreement with linear theory.  This raises the
question as to what the appropriate comparison smoothing scale should
be. The average distance at $z$=19 of the particle farthest from the
eventual halo centre was calculated to be 11.81, 11.23
and $10.64\,h^{-1}$ Mpc for cosmologies $\Omega_m$=0.25, 0.3, and 0.35
respectively.  Fig.~\ref{fig:hist_19} shows these values rather than
$R=10\,h^{-1}$ Mpc which would reflect larger values of the two-point
function at zero lag (by approximately 5 per cent). (Larger values of the
smoothing scale suppress the linear two-point functions more at
short-distances.)

\subsubsection{$\Psi$ as functions of redshift and cosmology}

\begin{figure}
\plotone{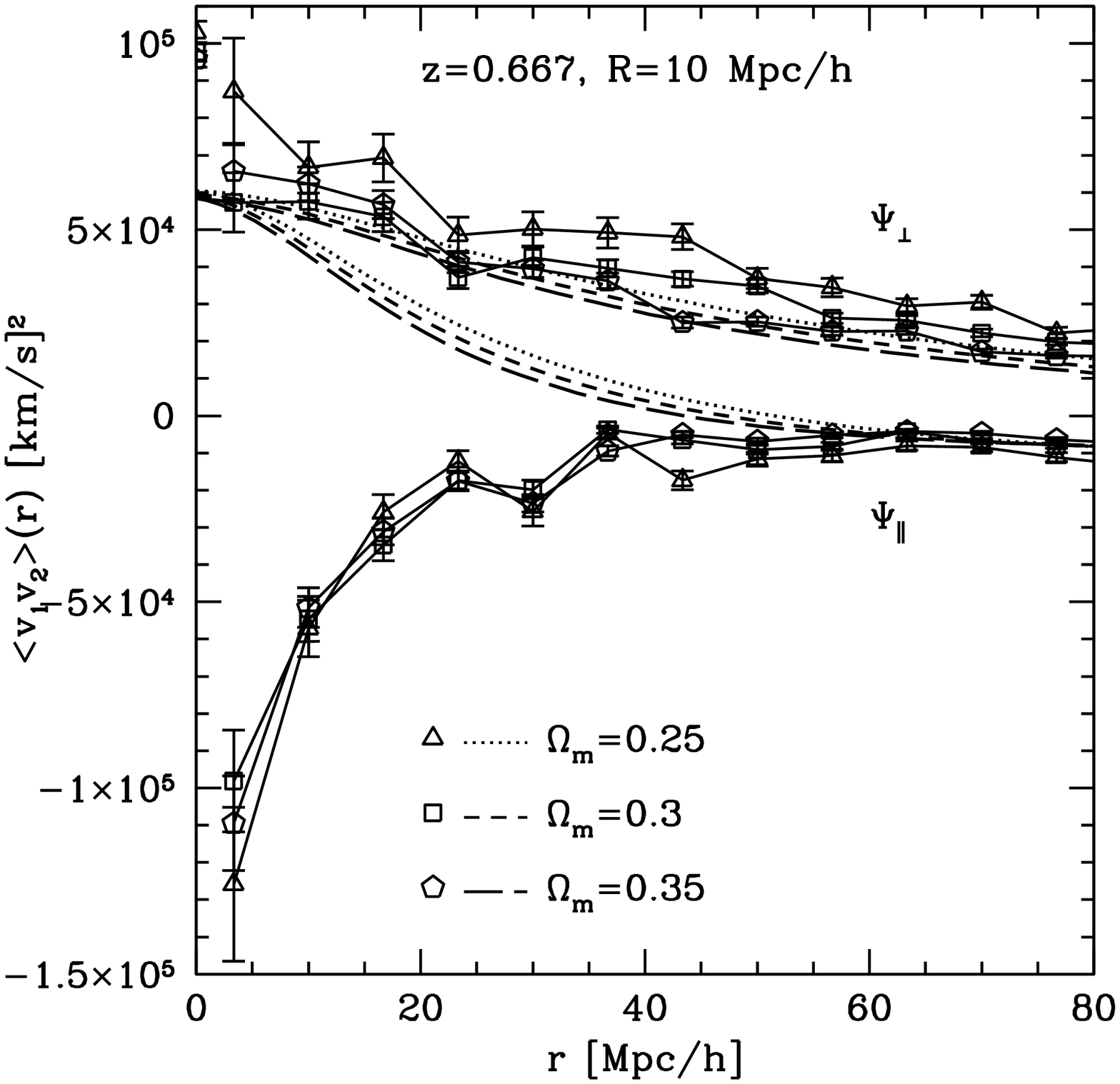}
\caption{as above, but for $z=0.667$.
 \label{fig:corr_0.667} }
\end{figure}

Examining the two-point functions for different $\Omega_m$ values at
different redshifts, four crucial results stand out.
First, looking at Table 1, it is essentially impossible to
discriminate between these three different cosmologies at zero lag at
redshift $z$=0, in contrast to the behaviour predicted by linear
theory \citep{peel02a}.  Second, at higher redshifts, the \rms
peculiar velocity dependence on $\Omega_m$ for these three values is
exactly opposite to that expected from linear theory (see 
Fig.~\ref{fig:theoryA}).  Third, the behaviour of the parallel component
shows heavy influence from infall for $r<30\,h^{-1}$ Mpc at {\em any}
redshift for which these massive haloes exist.  In particular, the
extreme anticorrelation seen in Fig.~\ref{fig:corr_0.667} at $r\sim
4\,h^{-1}$ Mpc  is comparable (or larger) in magnitude to the zero lag value
shown. Fourth, for the three values of $\Omega_m$ the perpendicular
components alone seem to  reflect linear theory closely in behaviour
if not in amplitude.  


\subsubsection{$\Psi$ as functions of history}

Tracking the history of the particles which will assemble the
largest haloes by $z$=0 (as well as those at higher redshift), 
it is clear that the parallel component shown
in Figs.~\ref{fig:hist_0.25} and \ref{fig:hist_0.667} does not
reflect the infall as strongly as seen in Figs.~\ref{fig:corr_0.25}
and \ref{fig:corr_0.667}.  Note also that the parallel
component shows remarkably little difference in behaviour (apart from
amplitude) between $z$=19 and $z$=0.667.  Prior to the assembly
of the largest haloes, the gross behaviour of their
particles is nearly fixed to track linear theory.  Only just as and
after a halo virializes does the strong infall out to $r<30\,h^{-1}$ Mpc
become apparent.  This is discussed in \S 5 as a coincidence in timing.

In Fig.~\ref{fig:hist_19}, the \rms
peculiar velocity is {\em smaller} than linear theory
predictions (and would be even smaller if we had retained $R=10\,h^{-1}$
Mpc smoothing).  This suppression is comparable to the excursion
heirarchy suppression proposed by \cite{BBKS} mentioned in \S 2.2
above.  The jump between the perpendicular component at
$r\gtrsim 2\,h^{-1}$ Mpc and the \rms value at $r$=0 is just as abrupt as
it is at later redshifts.  Some of this may be explained as an
artefact of resolution (recall the average interparticle separation is
$3\,h^{-1}$ Mpc).  Nonetheless, the perpendicular component
for the particles destined for large haloes is actually less like
linear theory here then at any subsequent time.

\section{Discussion}

There are at least three different factors which help explain the
differences between cluster-sized dark matter halo peculiar velocity
two-point functions at different values of $\Omega_m$ and their
differences from the two-point functions predicted by 
linear theory.  Specifically, there are two selection biases to
consider as well as the effect of dark energy domination.  
The first selection bias (also considered in previous work) is that 
regions which harbor the seeds of large dark matter haloes are
by definition overdense.  This helps explain why the \rms peculiar
velocity is 30 per cent higher than predicted by linear theory at any
redshift in which one would find such a virialized large dark matter
halo.

The second, more subtle selection bias strongly affects the
constraining power of a galaxy cluster-based set of velocity observations.
The rarity of large haloes is sensitive to a combination of $\Omega_m$
and $\sigma_8$.  If one fixes the value of $\sigma_8$ but decreases
the value of $\Omega_m$, there is an effective transfer of power from
small scales to large ones.  This creates deeper initial
potential wells at the
(now rarer) largest scales and thereby increases the acceleration of
halo velocities.  The largest mass haloes will preferentially be found
near these large rare fluctuations.  If we could calculate the velocity
statistics using {\em all} haloes
down to galaxy size (many of which are not responding to the extremely
rare, deep, large-scale potentials), we would find that there was a
decrease in the velocity power at small scales and would recover the expected
$\Omega_m$ dependence predicted by linear theory. However, the hot
plasma necessary to produce a
Sunyaev--Zel'dovich effect observation requires a 
deep potential well and is therefore sensitive to a specific halo mass
cutoff. 
This is why the \rms peculiar velocities of the $\Omega_m$=0.25 haloes
{\em for a high mass cutoff} were comparable or in excess of those for
larger values of $\Omega_m$ even at higher redshifts.  The very slight
mass dependence (which seems statistically insignificant in 
Fig.~\ref{fig:sigmav_vs_M}) is exactly the issue when varying $\Omega_m$.
It is clear that if we were to use an ever larger mass cutoff in that
figure, the average \rms velocity would {\em increase} while the
number of haloes used decreased.

The anticorrelation in the parallel component is simply a result of
infall, a non-linear process to which the linear theory is predictably
blind.  This behaviour was noted in \cite{croft95}; in an effort to
continue using linear predictions, the authors adjusted the smoothing
length in the linear model and selected velocities below a cutoff.
They conclude this approach introduces a strong bias in parameter
determination.  We examined this effect at different redshifts to
determine how this discrepancy with linear theory has evolved.
By redshift $z$=0.667, the few haloes massive enough to
produce a Sunyaev-Zel'dovich effect are so rare as to have come from
extremely overdense regions where collapse has been accelerated as if
from a much higher $\Omega_m$ universe.  Consequently, infall is
apparent out to large separations ($\sim 40\,h^{-1}$ Mpc).  
However, at lower redshifts 
where the rarity of such regions decreases, the anticorrelation scale
and its overall (absolute) magnitude also decrease
because the simulation is beginning to enter the era of dark
energy domination (simply represented in our work by a cosmological constant).
The extreme anticorrelation seen in Fig.~\ref{fig:corr_0.667} gives
way to a much milder anticorrelation only seen at the smallest
separations by $z$=0.  Dark matter--cosmological constant
equality occurs at $z$=0.44, 0.33 and 0.23 for $\Omega_m$=0.25, 0.3
and 0.35 respectively and 
marks the beginning of the end for new infall.  Large modes which
have not yet collapsed begin to
decay during the acceleration phase, so volumes not undergoing
gravitational collapse never will.  

Fig.~\ref{fig:corr_0.0} demonstrates that 
the cosmological constant has become dominant more recently for the larger
$\Omega_m$ values because the parallel anticorrelation at $r\sim 2.5\,h^{-1}$
Mpc is still comparable (though negative) to about half the \rms velocity.
In contrast, for $\Omega_m$=0.25, the growth of structure at
these comoving scales has already halted and the expansion has accelerated
sufficiently to begin a decay towards linear theory values, especially
for $r\gtrsim 5\,h^{-1}$ Mpc.  

Finally, we remark on the history of particles destined to be in
large mass haloes by $z$=0.
The \rms peculiar velocity derived from these sets of particles at $z$=19
(Fig.~\ref{fig:hist_19}) is lower than the linear theory
prediction  in agreement with the peak-background split prediction of
\cite{BBKS}.  In addition, the perpendicular and parallel components 
for separations $r\lesssim 40\,h^{-1}$ Mpc are also lower in amplitude
than linear theory predictions.  However, as the haloes assemble in
the heirarchical paradigm (Figs. \ref{fig:hist_0.25} and
\ref{fig:hist_0.667}), the \rms peculiar velocity surpasses linear
theory and the perpendicular components (from $2\lesssim r \lesssim
40\,h^{-1}$ Mpc) evolve towards linear theory values.  The perpendicular
component represents pairs of haloes responding only to some third 
large-scale fluctuation.  It therefore probes the field more
effectively than the parallel component in which the pairs' 
self-attraction at these scales dominates (on average) over gravitational
attractions from other potential wells.  Consequently, the
perpendicular component gradually recovers the behaviour predicted by
linear theory, modulo an overall boost in amplitude created by the
mass selection bias mentioned above.  In contrast, the
parallel component gradually reflects the ever more common large scale
infall until matter domination ends.  The effect is heuristically like
the function $\Phi_\para$ mentioned in \S 2 above, though the scale
occurs at smaller separation and the effect has a larger amplitude
due to the non-linearity of the local density.

By $z$=0.25, when
many of the $z$=0 haloes have more than 60 particles within their
virial radii, the parallel components
begin to display the anticorrelation more apparent in
Figs.~\ref{fig:corr_0.0}--\ref{fig:corr_0.667}.  The coincidence in
timing mentioned in \S 4.3.2 refers to the rarity of large scale
potential wells compared to their amplitudes.  As the largest mass haloes
become more common, the infall felt by the ones still forming comes
from more common smaller
amplitude potential wells; consequently, there is less apparent infall
in Fig.~\ref{fig:hist_0.667} than Fig.~\ref{fig:corr_0.667} simply
because the volume probed by the particles destined for haloes is much larger
than the rare, high-amplitude fluctuation dominated volume probed by
fully formed high mass haloes.

\section{Conclusions}

The purpose of this work was two-fold.  The primary goal was to
examine how $\Omega_m$ alone affects the velocity statistics of
cluster-sized haloes using simulations as opposed to what linear
theory predicts.  The secondary purpose was to show that small
($128^3$ particle) N-body simulations are sufficient to characterize
the statistics of cluster-sized dark matter haloes and
furthermore are fast enough to allow an exploration of parameter space
in a reasonable amount of time.  

We have shown that the dependence on $\Omega_m$ for the \rms peculiar velocity
for a fixed value of $\sigma_8$ and a fixed lower cutoff for the halo 
mass is counterintuitive to what linear theory predicts with a fixed
$\sigma_8$.  The peculiar velocities which develop for the
largest mass haloes are faster for a smaller value of $\Omega_m$ while
the universe is matter dominated because the largest modes must have a
greater amplitude for a fixed value of $\sigma_8$.  
However, since the decay of the (uncollapsed)
largest modes occurs earlier for a smaller $\Omega_m$, there is a
convergence towards an indistinguishable \rms peculiar velocity at
$z\sim$0 even though
(in our work) $\Omega_m$ varied by $\sim16$ per cent.
Peculiar velocity catalogs from  kinetic Sunyaev-Zel'dovich
observations therefore have a much more complicated dependence
on $\Omega_m$ than
other data from, \eg, the Cosmic Microwave Background, or large-scale
galaxy redshift surveys.

In addition, we have shown that the two-point velocity functions 
(in real space) for cluster-sized dark matter haloes are
as sensitive to these complicating biases as the \rms peculiar velocity
alone.  The prospect of using the information from the two-point
functions themselves to constrain $\Omega_m$ are currently hindered by two
major observational factors.  First, the 
mass-based selection bias mentioned above affects the perpendicular
component in a manner very similar to its effect on the \rms
velocity.  Second, the sources of noise mentioned in \S 1
are not
guaranteed to be stochastic, yet are comparable in amplitude to the
\rms signal ($\approx 200\ \kms$)\citep{knoxchurch, nagai03}.

The full cosmological sensitivity of a kinetic Sunyaev--Zel'dovich
derived peculiar velocity catalog has not been explored.  This is
where our secondary goal will come into play.
Rough scaling shows the potential for a reasonable search through
parameter space.  For example, 
based on cluster statistics using multiple small 
($128^3$ particle) ART simulations on a current, 
fairly large supercomputer or multiple efficient clusters, we find:
\bea
\label{eqn:time}
T & \sim & \left(N_{\mathrm{parameters}}\over 5\right) 
        \left(N_{\mathrm{values/parameter}}\over 3\right) \\
& &     \left(N_{\mathrm{realizations/value}}\over 10\right)
        \left(8_{\mathrm{CPU/realization}}\over N_{\mathrm{CPUs}}\right)
 150\ {\mathrm{weeks}}. \nonumber
\eea
For instance, five parameters at three values each with ten
realizations using only 160 CPUs would run in about two months.
While the rest of Eq.~(\ref{eqn:time}) does scale linearly, 
the number of CPUs per realization is a non-linear function of the
architecture of the computer and the performance of the simulation.
The choice of $8_{\mathrm{CPU/realization}}$ is based on our simulations for
this project  and is only given as a guideline.
Considering Moore's law, this an ever more conservative estimate.  This is
not on par with the speed of a Monte--Carlo Markov Chain method.  However,
when simulations are the only reliably accurate way to explore
parameter space, a simpler Fisher Matrix type approach is now 
conceivable for multiple parameters.

\section*{Acknowledgments}
AP would like to thank I.~Zehavi, L.~Knox, G.~Efstathiou, N.~Turok,
A.~Kravtsov and R.~Sheth 
for useful conversations; in particular, AP would like to
especially thank D.~Nagai for his assistance in running simulations.
AP is grateful for the anonymous referee's comments, which pointed out
how to strengthen the final version of this paper.
AP was supported at DAMTP by a PPARC rolling grant and at 
UC Davis by NASA GSRP Fellowship No.~1564 and NSF grant
No.~0307961.  The initial work in \S 3.4 was conducted on the IBM SP
parallel computer `Seaborg' at the National
Research Scientific Computing Center in Berkeley, CA which is
supported by the Office of Science of the Department of Energy under
contract No.~DE-AC03-76SF00098.  The multiple-cosmology aspect of
this work was finished using the UK National Cosmology Supercomputer
in Cambridge, UK under a PPARC Rolling Grant.

\section*{Appendix: Two-Point Correlations at Peaks}

This appendix is provided primarily as a pedagogical tool.  The nonlinear
evolution of clusters and the mass selection effect of kSZ
observations renders the consideration of peak \rm{vs.~}background
statistics moot here, especially since, as shown below and in past
papers, the peak statistics predict suppressed velocities compared to
background, \ie, general field velocities.  However, future methods of
observation and statistical analysis not mentioned in this paper may find
the calculations presented below quite relevant.

We parallel the approach detailed in \citet{BBKS} for calculating the
peak statistics for a Gaussian distributed linear density field,
$\delta(\vecx)$.  The density gradients at the peak, $\eta_i(\vecx_p)
= \del_i\delta|_{x_p}$, are zero (and nearly zero in some peak
neighborhood) and the second derivatives of the density at the peak,
$\zeta_{ij}(\vecx_p) = \del_i\del_j\delta|_{x_p}$, 
form a $3\times3$ symmetric matrix
with positive eigenvalues.  The six independent entries are relabelled
$\zeta_A$ where $\{A=1,2,3\}$ correspond to the diagonal elements
(originally $\zeta_{11}, \zeta_{22}$ and $\zeta_{33}$) and
$\{A=4,5,6\}$ correspond to the off-diagonals (originally $\zeta_{23},
\zeta_{13}$ and $\zeta_{12}$).

We shall adopt much of the notation found in \citet{BBKS}.  We replace
$\zeta_1, \zeta_2, \zeta_3$, with $x,y,z$ where:
\bea
x & = & -(\zeta_1+\zeta_2+\zeta_3)/\sigma_2\nonumber \\
y & = & -(\zeta_1-\zeta_3)/(2\sigma_2)\\
z & = & -(\zeta_1-2\zeta_2+\zeta_3)/(2\sigma_2)\nonumber\ .
\eea
Then, $\delta$ correlates only with $x$, not $y$, $z$, $\veceta$, 
nor $\zeta_{4,5,6}$.  We also scale $\delta$ by $\sigma_0$,
and thus have constructed unitless variables with simple Gaussian
widths: $\langle \nu^2 = \delta^2/\sigma_0^2\rangle = \langle x^2
\rangle = 15\langle y^2 \rangle = 5\langle z^2 \rangle = 1$.  In
addition, if we ignore the velocities, we have a nearly symmetric
auto-correlation matrix: the only non-zero, independent off-diagonal
element is $\langle \nu x\rangle = \sigma_1^2/(\sigma_0^2\sigma_2^2)$.

Of course, we are also concerned with the velocities.  In linear
perturbation theory, the divergence of the velocity field is given by
the time derivative of the density field (Eq.~(\ref{eqn:continuity})).
Therefore, each component $v_i$ only correlates with $\eta_i$, and
to estimate the peak \rms velocity we only need to examine the
$6\times6$ element matrix:
\be
M=\left(\begin{array}{c c}\langle \vecv \vecv \rangle & \langle \vecv
  \veceta \rangle \\ \langle \veceta \vecv \rangle & \langle \veceta
  \veceta \rangle \end{array}\right) = {1\over 3}\left(\begin{array}{c
    c} \sigma_v^2\ \mathrm{I}_3 & \dot{D}\sigma_0^2\ \mathrm{I}_3 \\
  \dot{D}\sigma_0^2 \ \mathrm{I}_3 & \sigma_1^2\
  \mathrm{I}_3\end{array} \right).
\ee
($\mathrm{I}_3$ is simply the $3\times 3$ identity matrix).  Inverting
this matrix and looking at the quadratic form $Q = \vecd^T M^{-1}
\vecd/2$ which appears in the Gaussian pdf $P \propto e^{-Q}$
($\vecd=[v_i,\eta_i]$), while imposing the condition that
$\veceta(\vecx_p)=0$ leads to the expression given in
Eq.~(\ref{eqn:peakstat}) above for the autocorrelations of {\em peak}
velocities.

Complicating matters, however, is the fact that many of the
corresponding two-point correlations are not zero.  One must reexamine
the entire correlation matrix to see the effects on the peak velocity
two-point functions (or, in fact, any other two-point functions).  The
list of relevant variables is now $\vecd = [\delta_1, \vecv_1,
      \veceta_1, x_1, y_1, z_1, \zeta_{1A}, \delta_2, \vecv_2,
      \veceta_2, x_2, y_2, z_2, \zeta_{2A};\ A=4,5,6]$, 
where the subscripts
$\{1,2\}$ refer to the points $\vecx$ and $\vecx + \vecr$,
respectively.  Our correlation matrix has now blossomed to $26\times
26$ elements\footnote{``Even for the two-point function, the task of
  integrating over all these variables is not pleasant to
  contemplate.'' \citep{BBKS}}.

If we continue to follow \citet{BBKS}, the next step would be to
rotate $\zeta$ onto its principle axes.  However, we cannot rotate
both $\zeta_1$ and $\zeta_2$ simultaneously. Instead, it makes more
sense to rotate the $\hat{\vecz}$ axis to be in alignment with the
line-of-sight between the two points, \ie, along $\vecr$.  This is
what is done in evaluating Eq.~(\ref{eqn:psitheory}).

The full analytic forms for the linear perturbation theory velocity
correlation functions are often given as:
\bea
\label{eqn:finalperp}
\Psi_\perp^{-1} &=& {\dot{D}_1\dot{D}_2\over 2\pi^2} \int dk
|\delta_{k,0}|^2 {j_1(kr)\over kr}W^2(kR)\\
&=&{\dot{D}_1\dot{D}_2\over 3} \int {k^2dk\over 2\pi^2} 
{|\delta_{k,0}|^2\over k^2}[j_0(kr) + j_2(kr)]W^2(kR)\nonumber
\eea
and:
\bea
\label{eqn:finalpara}
\Psi_\para^{-1} &= {\dot{D}_1\dot{D}_2\over 2\pi^2} \int dk
|\delta_{k,0}|^2 [j_0(kr) - 2j_1(kr)/kr]W^2(kR)\\
&= {\dot{D}_1\dot{D}_2\over 3} \int {k^2dk\over 2\pi^2}
{|\delta_{k,0}|^2\over k^2}[j_0(kr) - 2j_2(kr)]W^2(kR)\nonumber
\eea
where we have explicity added a $-1$ superscript: in general, both the
parallel and perpendicular
$\Psi^{n}$s will be defined as above, with a $k^{2n}$ in the integral,
much like the role of the $n$ subscript for Eq.~(\ref{eqn:sigman}).
We also define:
\be
\phi_m^n(r) =  D_1D_2\int {k^2dk\over 2\pi^2}
|\delta_{k,0}|^2 j_m(kr)\ k^n\ W^2(kR).
\ee
Below, we will implicitly assume that all $\Psi$, $\phi$ and
$\sigma$ functions are in terms of growth functions, not their
time derivatives; thus we will use appropriate factors of
$\beta=\dot{D}/D$ whenever velocities appear.

The second lines in Eqs.~(\ref{eqn:finalperp}) and
(\ref{eqn:finalpara}) make it clear that we can rewrite
$\Psi_\perp^{n}$ above as $\beta_1\beta_2 (\phi^{2n}_0 +
\phi^{2n}_2)/3$ and similarly $\Psi_\para^{n} = \beta_1\beta_2
(\phi^{2n}_0 - 2\phi^{2n}_2)/3$.  Also note that $\xi(r) =
\phi^0_0(r)$ and $\sigma_n^2 = \phi^{2n}_0(0)$.  In this way, all the
auto- and two-point correlations are expressible as linear
combinations of the set of functions $\phi_m^n(r)$.  As mentioned
above, we will always rotate the $z$-axis to be parallel to
$\hat{\vecr}$.

We solve for and list the various auto- and two-point correlations
below, with subscript $a=1,2$ for the two points $\vecx$ and $\vecx +
\vecr$ respectively.  For clarity, we also label the Kronecker-delta
function with a superscript $K$.  For completeness, the
auto-correlations are repeated here:
\bea
\langle \nu_a\nu_a \rangle &=& 1\nonumber\\
\langle v_{ai} v_{aj}\rangle & = & \delta_{ij}^K\beta_a^2
\sigma_{-1}^2 /3\nonumber\\
\langle v_{ai} \eta_{aj}\rangle & = & \delta_{ij}^K\beta_a \sigma_0^2
/3\nonumber\\
\langle \eta_{ai} \eta_{aj}\rangle & = & \delta_{ij}^K\sigma_1^2 
/3\nonumber\\
\langle \nu_a x_{a} \rangle &=&
\sigma_1^2/(\sigma_0\sigma_2)\nonumber\\
\langle x_a x_{a} \rangle &=& 1;\ \langle y_a y_{a}
\rangle=1/15; \ \langle z_a z_{a} \rangle=1/5\nonumber\\
\langle \zeta_{ai}\zeta_{aj} \rangle
& = & \delta_{ij}^K\sigma_2^2/15\nonumber\\
& & \{i,j=4,5,6\}\nonumber
\eea
all others being zero.  For the following, we suppress the $r$
argument as given, and rotate $\hat{\vecz} \para \hat{\vecr}$.  The
two-point correlations are:
\bea
\langle \nu_1\nu_2 \rangle &=& \phi^0_0/\sigma_0^2\nonumber \\
\langle \nu_1v_{2i} \rangle &=& -\delta_{i3}^K\beta_2
\phi^{-1}_1/\sigma_0\nonumber\\
\langle \nu_2v_{1i} \rangle &=& \delta_{i3}^K\beta_1
\phi^{-1}_1/\sigma_0\label{eqn:v1tov2}\\
\langle \nu_1\eta_{2i} \rangle &=& -\delta_{i3}^K
\phi^1_1/\sigma_0 = -\langle \nu_2\eta_{1i}\rangle\nonumber \\
\langle \nu_1 x_2\rangle & = & \phi^2_0
/(\sigma_0\sigma_2)\nonumber\\
\langle \nu_1 y_2\rangle & = & \phi^2_2
/(2\sigma_0\sigma_2)\nonumber\\
\langle \nu_1 z_2\rangle & = & -\phi^2_2
/(2\sigma_0\sigma_2)\nonumber\\
\langle \nu_1 \zeta_{2i}\rangle & = & 0; \ \ \ \ \
\{i=4,5,6\}\nonumber\\
\langle \vecv_1 \vecv_2\rangle & = & \beta_1\beta_2
[\Psi_\perp^{-1}{{\mathbf I}}
+ (\Psi_\para^{-1} -
\Psi_\perp^{-1})\hat{\vecr}\hat{\vecr}]\nonumber\\
\langle \vecv_1 \veceta_2\rangle & = & \beta_1
[\Psi_\perp^0{{\mathbf I}} 
+ (\Psi_\para^0 - \Psi_\perp^0)\hat{\vecr}\hat{\vecr}]\nonumber\\
\langle \vecv_2 \veceta_1\rangle & = & \beta_2
[\Psi_\perp^0\hat{{\mathbf I}} 
+ (\Psi_\para^0 -
\Psi_\perp^0)\hat{\vecr}\hat{\vecr}]\label{eqn:v1tov2b}\\
\langle v_{1i} x_2\rangle & = &
-\delta_{i3}^K\beta_1\phi^1_1/\sigma_2 \nonumber\\
\langle v_{1i} y_2\rangle & = &
\delta_{i3}^K\beta_1(\phi^1_1/5-3\phi^1_3/10)/\sigma_2 \nonumber\\
\langle v_{1i} z_2\rangle & = & - \langle v_{1i} y_2\rangle\nonumber\\
\langle v_{1i} \zeta_{2j}\rangle & = &
\beta_1(\phi^1_1+\phi^1_3)(\delta^K_{i2}\delta^K_{j4} +
\delta^K_{i1}\delta^K_{j5})/5  \ \ \ \{j=4,5,6\}\nonumber\\
\langle \veceta_1 \veceta_2\rangle & = & [\Psi_\perp^1\hat
{{\mathbf I}}
+ (\Psi_\para^1 - \Psi_\perp^1)\hat{\vecr}\hat{\vecr}]\nonumber\\
\langle \eta_{1i} x_2\rangle & = &
-\delta_{i3}^K\phi^3_1/\sigma_2 = -\langle \eta_{2i}
x_1\rangle\label{eqn:eta1toeta2}\\
\langle \eta_{1i} y_2\rangle & = &
\delta_{i3}^K(\phi^3_1/5-3\phi^3_3/10)/\sigma_2 = -\langle \eta_{2i}
y_1\rangle\nonumber\\
\langle \eta_{1i} z_2\rangle & = & - \langle \eta_{1i} y_2\rangle 
= -\langle \eta_{2i} z_1\rangle\nonumber\\
\langle \eta_{1i} \zeta_{2j}\rangle & = &
(\phi^3_1+\phi^3_3)(\delta^K_{i2}\delta^K_{j4} +
\delta^K_{i1}\delta^K_{j5})/5  \ \ \ \{j=4,5,6\}\nonumber\\
\langle x_1x_2\rangle &=& \phi^4_0/\sigma_2^2\nonumber\\
\langle x_1y_2\rangle &=& \phi^4_2/(2\sigma_2^2)
= -\langle x_1z_2\rangle \nonumber\\
\langle y_1y_2\rangle &=& [19\phi^4_4/140 - \phi^4_2/21 +
  \phi^4_0/15]/ \sigma_2^2\nonumber\\
\langle y_1z_2\rangle &=& (\phi^4_2/7 - 3\phi^4_4/28)/
\sigma_2^2\nonumber\\
\langle z_1z_2\rangle &=& [27\phi^4_4/140 + \phi^4_2/7 +
  \phi^4_0/5]/\sigma_2^2\nonumber\\
\langle \zeta_{14}\zeta_{24}\rangle &=&
[-4\phi^4_4/35-\phi^4_2/21 + \phi^4_0/15]/\sigma_2^2\nonumber\\
& = & \langle \zeta_{15}\zeta_{25}\rangle \nonumber\\
\langle \zeta_{16}\zeta_{26}\rangle &=&
[\phi^4_4/35+2\phi^4_2/21 + \phi^4_0/15]/\sigma_2^2\nonumber
\eea
Note that most of the correlations are the same under
$\vecr \rightarrow -\vecr$ except those involving one $\veceta$ or one
$\vecv$ which pick up a minus sign \eg, Eqs.~(\ref{eqn:v1tov2}) and
(\ref{eqn:eta1toeta2}); and those involving only one $\vecv$ which
forces $D_1\rightarrow D_2$ and $\dot{D}_2\rightarrow\dot{D}_1$, \eg,
Eqs.~(\ref{eqn:v1tov2}) and (\ref{eqn:v1tov2b}).  In the limit where
we ignore correlations at distances large enough to reflect a
significant difference between $D_1$ and $D_2$ or between their time
derivatives, this becomes a nitpicking concern.

To understand peak velocity two-point correlations, we must now
perform the following steps.  First, we construct the required
$26 \times 26$ matrix $M$ using the above results and invert it.  
Next, we integrate the probability distribution given by $P \propto
e^{-Q}$ where $Q = \vecd^TM^{-1}\vecd/2$ over all the variables except
the velocities: over $\nu_1$ and $\nu_2$ from the peak cutoff (such as
$\approx 3$) to infinity; and over all values of the second derivative
which meet the positive eigenvalue criterion mentioned above.  Third,
we set the density derivatives $\eta_{ai}$ equal to zero.  The last
step is to read off the remainder of the correlation matrix for the
velocities, much like what was done to derive Eq.~(\ref{eqn:peakstat})
above.  

\bibliography{vpec}
\label{lastpage}
\end{document}